\documentclass[aps,twocolumn,floatfix,groupedaddress,nofootinbib]{revtex4}
\usepackage{csquotes}
\usepackage{graphicx,color}
\usepackage{float}
\usepackage[caption=false]{subfig}
\usepackage{amsmath}
\usepackage{amssymb}
\usepackage{multirow}
\usepackage{dsfont}
\expandafter\let\csname equation*\endcsname\relax
\expandafter\let\csname endequation*\endcsname\relax

\begin{document}

\title{Quantum version of the integral equation theory based dielectric scheme for strongly coupled electron liquids\vspace*{-1.45mm}}
\author{Panagiotis Tolias$^{1}\footnotemark\footnotetext{\vspace*{-2.40mm}corresponding author: tolias@kth.se\vspace*{-10.40mm}}$, Federico Lucco Castello$^{1}$ and Tobias Dornheim$^{2,3}$}
\affiliation{$^1$ Space and Plasma Physics - Royal Institute of Technology (KTH), SE-10044 Stockholm, Sweden\\
             $^2$ Center for Advanced Systems Understanding (CASUS), D-02826 G\"orlitz, Germany\\
             $^3$ Helmholtz-Zentrum Dresden-Rossendorf (HZDR), D-01328 Dresden, Germany\vspace*{-1.45mm}}
\begin{abstract}
\noindent A novel dielectric scheme is proposed for strongly coupled electron liquids that handles quantum mechanical effects beyond the random phase approximation level and treats electronic correlations within the integral equation theory of classical liquids. The self-consistent scheme features a complicated dynamic local field correction functional and its formulation is guided by \emph{ab initio} path integral Monte Carlo simulations. Remarkably, our scheme is capable to provide unprecedently accurate results for the static structure factor with the exception of the Wigner crystallization vicinity, despite the absence of adjustable or empirical parameters.
\end{abstract}
\maketitle

\noindent The three dimensional uniform electron fluid (UEF) constitutes the simplest, yet realistic, bulk electronic system capable of exhibiting strong correlation effects\,\cite{GrossBo,MahanBo,quanele}. It is often referred to as homogeneous electron gas, jellium or quantum one-component plasma being the quantum analogue of the classical one-component plasma (OCP)\,\cite{cOCPrev,IchiRMP}. Early studies focused on the ground state at metallic densities due to its relevance for valence electrons in simple metals\,\cite{MahanBo,KittelB}. These investigations have led to remarkable insights such as the Landau Fermi-liquid theory\,\cite{FermiLi} and the Bohm \& Pines quasi-particle picture of collective excitations\,\cite{BohmPin}. In addition, the accurate parametrization of different UEF properties based on systematic quantum Monte Carlo (QMC) simulations\,\cite{ground1,ground2,ground3} has been of vital importance for the unrivaled success of density functional theory in the description of real materials\,\cite{DFTKoSh,DFTJone}.

Increasing interest in warm dense matter (WDM)\,\cite{WDMint1,WDMint2,WDMbook}, which is an exotic yet ubiquitous state of high temperature highly compressed matter encountered in dense astrophysical objects\,\cite{WDMapp1,WDMapp2,WDMapp3}, fuel compression processes during inertial confinement fusion\,\cite{WDMapp4} and novel material fabrication\,\cite{WDMapp5}, has provided the impetus for a worldwide intense research activity targeted at the high density finite temperature UEF\,\cite{BrowPRL,DornRev,KaraPRB,BoniRev,DornPoP}. Numerous breakthroughs concerning the development of QMC methods or of QMC extensions that alleviate the fermion sign problem at different WDM regions\,\cite{newQMC1,newQMC2,newQMC3,newQMC4,newQMC5,newQMC6,newQMCa} as well as progress regarding the correction of finite-size errors\,\cite{newQMC7,newQMCb,newQMCc} and the numerical implementation or even full circumvention of analytic continuation\,\cite{newQMC8,newQMC9,newQMC0} have led to a very accurate description of the thermodynamic, static, dynamic and non linear behavior of the UEF in WDM conditions\,\cite{WDMsta1,WDMsta2,WDMsta3,WDMsta4,WDMsta5,WDMsta6,WDMsta7,WDMsta8}.

On the other hand, the physical realization of dilute ground state homogeneous electronic systems is known to be extremely challenging, while low density finite temperature inhomogeneous electron systems are inaccessible even to state-of-the-art experiments. Therefore, much less attention has been paid to the strongly coupled electron liquid regime (hence the use of the label UEF instead of the label jellium which typically concerns the gaseous phase) especially at finite temperatures, despite preliminary confirmations and legitimate speculations of exotic collective behavior that could be of technological importance. These include phenomena that begin to manifest themselves at the margin of the strongly coupled UEF regime. A prominent example concerns the onset of effective attraction between two electrons due to short range order\,\cite{introM1}, its manifestation onto the negative sign of the spin-offdiagonal component of the static density response function\,\cite{introM2} as well as its association with the possibility of Cooper pairing\,\cite{BCStheo} and thus the potential emergence of super-conductivity without phonons\,\cite{introM3,introM4}. Another closely related example concerns the existence of a minimum in the dispersion of density-density fluctuations and an associated roton feature in the dynamic structure factor\,\cite{newQMC8,newQMC9} as well as their microscopic finite temperature interpretation in terms of an electronic pair alignment model\,\cite{introM5} and their alternative ground state interpretation in terms of an excitonic mode\,\cite{introM6,introM7,introM8,introM9}. These also include phenomena that have been speculated to manifest themselves deep within the strongly coupled UEF regime such as the possibility of a charge-density wave instability\,\cite{introD0,introD1,introD2,introD3}, the emergence of a spin-density wave\,\cite{introD3,introD4,introD5,introD6} and the possibility of a continuous paramagnetic to ferromagnetic transition (Stoner's instability)\,\cite{introDS,introD7,introD8,introD9,introDN}.

The only exceptions to this rather discouraging state of affairs concern the thermodynamic and the static properties of the strongly coupled paramagnetic UEF that were recently characterized by extensive path integral Monte Carlo (PIMC) simulations at finite temperatures\,\cite{HNCPIMC,IETChem}. The highly-accurate results were compared to two novel schemes of the self-consistent dielectric formalism that are equipped to handle strong correlations. The hypernetted chain (HNC-) based scheme that handles quantum mechanical effects at the random phase approximation level and incorporates a frequency independent local field correction treating strong Coulomb correlations within the classical HNC approximation\,\cite{HNCSTLS,HNCPIMC}. The integral equation theory (IET-) based scheme that supplements the HNC-based scheme with a near-exact classical Coulomb bridge function\,\cite{IETChem,IETLett}. Both schemes were demonstrated to yield excellent predictions for the thermodynamic properties (benefitting from a favorable error cancellation), accurate predictions for the position of the static structure factor peak but quite inaccurate predictions for the magnitude of the static structure factor peak\,\cite{HNCPIMC,IETChem}. It is worth noting that interaction energy predictions of the consistently more accurate IET-based scheme were always within $0.68\%$ of the exact value\,\cite{IETChem}.

In the present communication, we substantially refine the treatment of quantum mechanical effects within the IET-based dielectric scheme guided by available PIMC simulations. The upgraded quantum version of the IET-based scheme features a complicated dynamic local field correction functional that leads to a substantially more involved set of equations. The associated numerical complexity proved to be rewarding, since our scheme yields excellent predictions for the static structure factor at all states except from the Wigner crystallization vicinity.

\emph{-- Uniform electron fluid}. The UEF is a homogeneous model system consisting of electrons immersed in a rigid ionic neutralizing background. We focus on the paramagnetic (unpolarized) case of equal spin-up and -down electrons, whose state points are fully specified by two dimensionless quantities: \textbf{(1)} the quantum coupling parameter $r_{\mathrm{s}}=d/a_{\mathrm{B}}$ with $d=(4\pi{n}/3)^{-1/3}$ the Wigner Seitz radius and $a_{\mathrm{B}}=\hbar^2/m_{\mathrm{e}}e^2$ the Bohr radius, \textbf{(2)} the degeneracy parameter $\Theta=T/E_{\mathrm{F}}$ with $T$ the temperature in energy units, $E_{\mathrm{F}}=\hbar^2k_{\mathrm{F}}^2/(2m_{\mathrm{e}})$ the Fermi energy and $k_{\mathrm{F}}=(3\pi^2n)^{1/3}$ the Fermi wavevector. The WDM regime is roughly demarcated by $0.1\lesssim{r}_{\mathrm{s}},\Theta\lesssim10$, while the strongly coupled regime corresponds to $r_{\mathrm{s}}\gtrsim20$, $\Theta\lesssim1$. The theoretical treatment of the UEF in the WDM and especially the strongly coupled regime is formidable due to the coexistence of quantum effects (exchange, diffraction), Coulomb correlations and thermal excitations, as reflected on the lack of small parameters\,\cite{DornRev,BoniRev,DornPoP}.

\emph{-- Self-consistent dielectric formalism}. It constitutes one of the most powerful and versatile microscopic formulations for the description of the dynamic, structural and thermodynamic properties of the UEF\,\cite{GeneDF1,GeneDF2,GeneDF3,GeneDF4,GeneDF5}. The dielectric formalism combines fundamental results of linear density response theory\,\cite{quanele} with an approximate closure stemming from the perturbative quantum kinetic theory of non-ideal gases\,\cite{quankin} or the non-perturbative integral equation theory of classical liquids\,\cite{IETliqu}. Particularly, in the polarization potential approach\,\cite{IchimaB} of linear response theory, the density response function $\chi(\boldsymbol{k},\omega)$ is expressed in terms of the ideal (Lindhard) density response $\chi_0(\boldsymbol{k},\omega)$ and the dynamic local field correction $G(\boldsymbol{k},\omega)$ (LFC) as
\begin{equation}
\chi(\boldsymbol{k},\omega)=\frac{\chi_0(\boldsymbol{k},\omega)}{1-U(\boldsymbol{k})\left[1-G(\boldsymbol{k},\omega)\right]\chi_0(\boldsymbol{k},\omega)}\,,\label{densityresponseDLFC}
\end{equation}
with $U(\boldsymbol{k})=4\pi{e}^2/k^2$ for the bare Coulomb interactions. Moreover, for the finite temperature UEF, the combination of the zero frequency moment sum rule and the quantum fluctuation--dissipation theorem (FDT) with the analytic continuation of $\chi(\boldsymbol{k},\omega)$ to the complex frequency plane $\omega+\imath0\to{z}$ lead to a static structure factor $S(\boldsymbol{k})$ (SSF) relation that involves the Matsubara summation
\begin{equation}
S(\boldsymbol{k})=-\frac{1}{{n}\beta}\displaystyle\sum_{l=-\infty}^{\infty}\widetilde{\chi}(\boldsymbol{k},\imath\omega_l)\,,\label{Matsubaraseries}
\end{equation}
with $\widetilde{\chi}(\boldsymbol{k},z)$ the analytically-continued density response function and $\omega_l=2\pi{l}/(\beta\hbar)$ the bosonic Matsubara frequencies\,\cite{TanIchi}. Finally, the self-consistent dynamic LFC, which describes Pauli exchange, quantum diffraction and Coulomb correlation effects beyond the random phase approximation (RPA), is given by a complicated SSF functional of the general form
\begin{equation}
G(\boldsymbol{k},\omega)\equiv{G}[S](\boldsymbol{k},\omega)\,.\label{functionalclosure}
\end{equation}
Eqs.(\ref{densityresponseDLFC},\ref{Matsubaraseries},\ref{functionalclosure}) constitute a set of non-linear functional equations to be solved for the SSF\,\cite{DornRev,TanIchi}. Multiple dielectric schemes have been developed that target various UEF phase diagram regions and only differ in the approximate treatment of the exact (unknown) LFC functional.

In the warm dense matter regime of weak-to-moderate coupling, it is worth to single out two dielectric schemes that have been empirically found to be very accurate. \textbf{(1)} The Singwi-Tosi-Land-Sj\"olander (STLS) scheme\,\cite{STLSgro,STLSfin}; a semi-classical scheme that includes quantum effects exclusively via the Lindhard density response and whose static LFC closure is derived by truncating the classical BBGKY hierarchy at its first member with the factorization ansatz $f_2(\boldsymbol{r},\boldsymbol{p},\boldsymbol{r}^{\prime},\boldsymbol{p}^{\prime};t)=f(\boldsymbol{r},\boldsymbol{p};t)f(\boldsymbol{r}^{\prime},\boldsymbol{p}^{\prime};t)g(\boldsymbol{r}-\boldsymbol{r}^{\prime})$ where $f_2$ is the two particle distribution function, $f$ is the single particle distribution function, $g$ is the  pair correlation function at thermodynamic equilibrium.\,\textbf{(2)} The quantum STLS (qSTLS) scheme\,\cite{qSTLSge,qSTLSgr,qSTLSFT}; a fully quantum scheme that treats exchange \& diffraction effects beyond the RPA level and whose dynamic LFC closure is derived by truncating the quantum BBGKY hierarchy within the Wigner representation at its first member with the same STLS ansatz. The respective LFC closures are given by
\begin{align*}
&G_{\mathrm{STLS}}(\boldsymbol{k})=-\frac{1}{n}\int\frac{d^3q}{(2\pi)^3}\frac{\boldsymbol{k}\cdot\boldsymbol{q}}{k^2}\frac{k^2}{q^2}[S(|\boldsymbol{k}-\boldsymbol{q}|)-1]\,,\\
&G_{\mathrm{qSTLS}}(\boldsymbol{k},\omega)=-\frac{1}{n}\int\frac{d^3q}{(2\pi)^3}\frac{\chi^0_{\boldsymbol{k},\boldsymbol{q},\omega}}{\chi^{0}_{\boldsymbol{k},\omega}}\frac{k^2}{q^2}[S(|\boldsymbol{k}-\boldsymbol{q}|)-1]\,,
\end{align*}
where $\chi^0_{\boldsymbol{k},\boldsymbol{k}^{\prime},\omega}$, with the property $\chi^0_{\boldsymbol{k},\boldsymbol{k},\omega}\equiv\chi^0_{\boldsymbol{k},\omega}$, is a three-argument ideal density response that is defined by
\begin{align*}
\chi_0(\boldsymbol{k},\boldsymbol{k}^{\prime},\omega)&=-\frac{2}{\hbar}\int\frac{d^3q}{(2\pi)^3}\frac{f_0\left(\boldsymbol{q}+\frac{1}{2}\boldsymbol{k}^{\prime}\right)-f_0\left(\boldsymbol{q}-\frac{1}{2}\boldsymbol{k}^{\prime}\right)}{\omega-\frac{\hbar}{m}\boldsymbol{k}\cdot\boldsymbol{q}+\imath0}\,.
\end{align*}

\begin{figure*}
	\centering
	\includegraphics[width=7.00in]{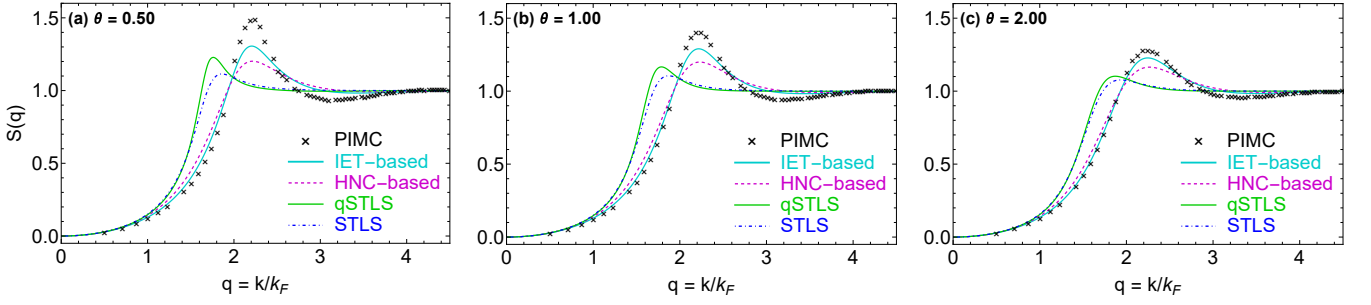}
	\caption{The paramagnetic electron liquid static structure factor for $r_{\mathrm{s}}=100$ and (a) $\Theta=0.5$,\,(b)\,$\Theta=1.0$,\,(c)\,$\Theta=2.0$. Results from the STLS scheme (dot-dashed blue line), qSTLS scheme (solid green line), HNC-based scheme (dashed magenta line), IET-based scheme (solid cyan line) and PIMC simulations (black crosses). The superiority of the IET-based scheme is apparent.}\label{fig:motivation}
\end{figure*}

In the liquid regime of moderate-to-strong coupling, it is worth to single out two very recent dielectric schemes. \textbf{(1)} The HNC-based scheme of Tanaka\,\cite{HNCSTLS,HNCPIMC}; a semi-classical scheme that treats quantum effects at the RPA level and combines the classical FDT, Ornstein-Zernike integral equation and HNC non-linear closure equation in order to derive a static LFC functional. \textbf{(2)} The IET-based scheme proposed by Tolias and coworkers\,\cite{IETChem,IETLett}; a semi-classical scheme that supplements the HNC-based scheme with a Coulomb bridge function, \emph{i.e.} it substitutes the approximate HNC closure equation with the exact IET closure equation benefitting from the recent extraction of classical OCP bridge functions from molecular dynamics simulations and their subsequent parametrization\,\cite{IETLett,OCPbrid,YOCPext}. The respective static LFC closures are
\begin{align*}
G_{\mathrm{HNC}}(\boldsymbol{k})&=-\frac{1}{n}\int\frac{d^3q}{(2\pi)^3}\frac{\boldsymbol{k}\cdot\boldsymbol{q}}{k^2}\frac{k^2}{q^2}\left[S(\boldsymbol{k}-\boldsymbol{q})-1\right]\nonumber\\&\quad\times\left\{1-\left[{G}(\boldsymbol{q})-1\right]\left[S(\boldsymbol{q})-1\right]\right\}\,.\nonumber\\
G_{\mathrm{IET}}(\boldsymbol{k})&=\frac{{B}(\boldsymbol{k})}{\beta{U}(\boldsymbol{k})}-\frac{1}{n}\int\frac{d^3q}{(2\pi)^3}\frac{\boldsymbol{k}\cdot\boldsymbol{q}}{k^2}\frac{k^2}{q^2}\left[S(|\boldsymbol{k}-\boldsymbol{q}|)-1\right]\nonumber\\&\quad\times\left\{-\frac{B(\boldsymbol{q})}{\beta{U}(\boldsymbol{q})}+1-\left[G(\boldsymbol{q})-1\right]\left[S(\boldsymbol{q})-1\right]\right\}\,,
\end{align*}
with $B(\boldsymbol{k})$ the Fourier transform of the bridge function $b(\boldsymbol{r})$. Note that, as expected, $G_{\mathrm{IET}}(\boldsymbol{k})\equiv{G}_{\mathrm{HNC}}(\boldsymbol{k})$ when $b(\boldsymbol{r})\equiv0$. The real space parametrization reads as\,\cite{IETLett,OCPbrid}
\begin{align}
&b(x,\Gamma)=\left[1-f(x)\right]b_{\mathrm{s}}(x,\Gamma)+f(x)b_{\mathrm{i}}(x,\Gamma)\,,\nonumber\\
&b_{\mathrm{s}}(x,\Gamma)=s_0(\Gamma)+\textstyle\sum_{i=2}^{5}s_i(\Gamma)x^{i}\,,\nonumber\\
&b_{\mathrm{i}}(x,\Gamma)=l_0(\Gamma)\Gamma^{5/6}\exp{\left[-l_1(\Gamma)(x-1.44)-0.3x^2\right]}\times\nonumber\\&\,\,\,\,\,\,\,\,\,\,\left\{\cos{\left[l_2(\Gamma)(x-1.44)\right]}+l_3(\Gamma)\exp{\left[-3.5(x-1.44)\right]}\right\},\nonumber\\
&f(x)=0.5\left\{1+\mathrm{erf}\left[5.0\left(x-1.5\right)\right]\right\}\,,\nonumber\\
&s_i(\Gamma)=\textstyle\sum_{j=0}^{3}s_{i}^{j}\Gamma(\ln{\Gamma})^{j}\,,\,\,\,l_i(\Gamma)=\textstyle\sum_{j=0}^{4}l_{i}^{j}\Gamma^{1/6}(\ln{\Gamma})^{j}\,,\nonumber
\end{align}
with $x=r/d$, $\Gamma=2\lambda^2(r_{\mathrm{s}}/\Theta)$ the classical coupling parameter, $\lambda^3=4/(9\pi)$ and $s_{i}^{j},l_{i}^{j}$ tabulated in Refs.\cite{IETLett,OCPbrid}.

\emph{-- QMC simulations in the liquid regime}. Extended and rigorously finite-size corrected (Refs.\cite{HNCPIMC,IETChem} for details), PIMC data are available for the interaction energy, static density response $\chi(\boldsymbol{k})\equiv\chi(\boldsymbol{k},\omega=0)$, static LFC $G(\boldsymbol{k})\equiv{G}(\boldsymbol{k},\omega=0)$ and SSF $S(\boldsymbol{k})$\,\cite{HNCPIMC,IETChem}. In particular, $34$ UEF states have been simulated in the phase diagram region defined by $20\leq{r}_{\mathrm{s}}\leq200,\,0.5\leq\Theta\leq{4}$. The boundary between the WDM and liquid regimes is of little interest for the present study, thus we focus on the $20$ UEF states that belong to the $50\leq{r}_{\mathrm{s}}\leq200,\,0.5\leq\Theta\leq{4}$ region.

Systematic comparison of the predictions of the STLS, qSTLS, HNC-based and IET-based schemes with the exact PIMC results led to the following conclusions\,\cite{HNCPIMC,IETChem}, see Fig.\ref{fig:motivation}: \textbf{(1)} The IET- \& HNC-based schemes yield very similar accurate predictions for the positions of the SSF and $\chi(\boldsymbol{k})$ extrema. The STLS and qSTLS schemes yield very similar predictions for the positions of the SSF and $\chi(\boldsymbol{k})$ extrema, but they significantly underestimate both. \textbf{(2)} The IET-based scheme result for the SSF and $\chi(\boldsymbol{k})$ peak magnitudes greatly improves the respective HNC-based result, but it still underestimates the exact PIMC result, mainly for the lowest $\Theta$ and highest $r_{\mathrm{s}}$ considered. The qSTLS scheme significantly improves the STLS result for the SSF and especially the $\chi(\boldsymbol{k})$ peak magnitudes, but the PIMC result underestimation persists. \textbf{(3)} The IET-based scheme consistently improves the HNC-based outcome for the $G(\boldsymbol{k})$ peak magnitude, though there is an underestimation compared to the exact PIMC outcome. On the other hand, the STLS \& the qSTLS schemes lead to static LFCs that do not feature a well-developed $G(\boldsymbol{k})$ maximum, but rather an extended plateau. \textbf{(4)} The IET- and HNC-based interaction energies are remarkably accurate owing to a favorable error cancellation (accuracies within $0.68\%$ and $1.37\%$, respectively).

\emph{-- Quantum version of the integral equation theory based scheme (qIET)}. In terms of physics, the main drawback of the IET-based scheme concerns the treatment of quantum effects on the RPA level. In terms of PIMC validation, the main drawback of the IET-based scheme concerns the underestimation of the SSF peak magnitude in spite of the accurate prediction of the SSF peak position. As discussed above, the qSTLS scheme improves the SSF peak magnitude prediction of the STLS scheme without affecting its SSF peak position prediction. Thus, physics and numerics grounds imply that an IET-level treatment of strong correlations combined with a qSTLS-level treatment of quantum effects could possibly alleviate the main deficiencies of the IET-based scheme. The dynamic LFC closure of our qIET-based scheme can be simply inferred by comparison.\,Since the qSTLS closure emerges from the STLS closure by a $(\boldsymbol{k}\cdot\boldsymbol{q})/k^2\to\chi^0_{\boldsymbol{k},\boldsymbol{q},\omega}/\chi^{0}_{\boldsymbol{k},\omega}$ substitution, the same substitution in the IET-based static LFC closure leads to the qIET-based dynamic LFC closure
\begin{align}
G&_{\mathrm{qIET}}(\boldsymbol{k},\omega)=\frac{{B}(\boldsymbol{k})}{\beta{U}(\boldsymbol{k})}-\frac{1}{n}\int\frac{d^3q}{(2\pi)^3}\frac{\chi^0_{\boldsymbol{k},\boldsymbol{q},\omega}}{\chi^{0}_{\boldsymbol{k},\omega}}\frac{k^2}{q^2}\left[S(|\boldsymbol{k}-\boldsymbol{q}|)-1\right]\nonumber\\&\,\,\quad\times\left\{-\frac{B(\boldsymbol{q})}{\beta{U}(\boldsymbol{q})}+1-\left[G(\boldsymbol{q},\omega)-1\right]\left[S(\boldsymbol{q})-1\right]\right\}.\label{qIETclosure}
\end{align}
Note the substitution ${G}(\boldsymbol{q})\to{G}(\boldsymbol{q},\omega)$ inside the integral. In an identical fashion, the dynamic LFC closure of the quantum version of the hypernetted chain based scheme (qHNC) can be constructed. It is given by
\begin{align}
G_{\mathrm{qHNC}}(\boldsymbol{k},\omega)&=-\frac{1}{n}\int\frac{d^3q}{(2\pi)^3}\frac{\chi^0_{\boldsymbol{k},\boldsymbol{q},\omega}}{\chi^{0}_{\boldsymbol{k},\omega}}\frac{k^2}{q^2}\left[S(\boldsymbol{k}-\boldsymbol{q})-1\right]\nonumber\\&\quad\times\left\{1-\left[{G}(\boldsymbol{q},\omega)-1\right]\left[S(\boldsymbol{q})-1\right]\right\}\,.\label{qHNCclosure}
\end{align}
The closed normalized set of equations for the qIET-based scheme comprises the normalization condition of the Fermi-Dirac energy distribution function [see Eq.(\ref{qIETfinal1})], the Fourier transform of the classical OCP bridge function [see Eq.(\ref{qIETfinal2})], the ideal Lindhard density response expressed through the auxiliary complex function $\Phi(x,l)$ and evaluated at the imaginary Matsubara frequencies $\omega_l$ including the static limit [see Eqs.(\ref{qIETfinal3},\ref{qIETfinal4})], the dynamic LFC triple integral expression expressed through the auxiliary complex function $\Psi(x,l)$ and evaluated at the imaginary Matsubara frequencies $\omega_l$ including the static limit [see Eqs.(\ref{qIETfinal5},\ref{qIETfinal6})] and the infinite Matsubara summation expression for the SSF [see Eq.(\ref{qIETfinal7})]. It is worth noting that Eqs.(\ref{qIETfinal5},\ref{qIETfinal6}) emerge from Eq.(\ref{qIETclosure}), after the introduction of azimuthally expanded two-center bipolar coordinates. In what follows, $\bar{\mu}=\mu/T$ denotes the normalized chemical potential, wavenumbers are expressed in $k/k_{\mathrm{F}}$ units and distances are expressed in $rk_{\mathrm{F}}$ units.
\begin{gather}
\int_0^{\infty}\frac{\sqrt{z}dz}{\exp{\left(z-\bar{\mu}\right)}+1}=\frac{2}{3}\Theta^{-3/2}\,,\,\label{qIETfinal1}\\
\frac{B(x)}{\beta{U}(x)}=\frac{9\pi}{8}\frac{\Theta}{r_{\mathrm{s}}}x\int_0^{\infty}yb\left(y,2\lambda^2\frac{r_{\mathrm{s}}}{\Theta}\right)\sin{\left(\frac{x}{\lambda}y\right)}dy\,,\label{qIETfinal2}
\end{gather}
\begin{widetext}
\begin{align}
&\Phi(x,l)=\frac{1}{2x}\int_0^{\infty}dy\frac{y}{\exp{\left(\frac{y^2}{\Theta}-\bar{\mu}\right)}+1}\ln{\left[\frac{\left(x^2+2xy\right)^2+\left(2\pi{l}{\Theta}\right)^2}{\left(x^2-2xy\right)^2+\left(2\pi{l}\Theta\right)^2}\right]}\,,\label{qIETfinal3}\\
&\Phi(x,0)=\frac{1}{\Theta{x}}\int_0^{\infty}dy\left[\left(y^2-\frac{x^2}{4}\right)\ln{\left|\frac{2y+x}{2y-x}\right|}+xy\right]\frac{y\exp{\left(\frac{y^2}{\Theta}-\bar{\mu}\right)}}{\left[\exp{\left(\frac{y^2}{\Theta}-\bar{\mu}\right)}+1\right]^2}\,,\label{qIETfinal4}\\
&\Psi(x,l)=-\frac{3}{8}\int_0^{\infty}\left\{\left[-\frac{B(u)}{\beta{U}(u)}+1\right]S(u)-\left[\frac{\Psi(u,l)}{\Phi(u,l)}\right]\left[S(u)-1\right]\right\}\frac{du}{u}\int_{|u-x|}^{u+x}w\left[S(w)-1\right]dw\nonumber\\&\qquad\quad\,\,\,\,\times\int_0^{\infty}\frac{ydy}{\exp{\left(\frac{y^2}{\Theta}-\bar{\mu}\right)}+1}\ln{\left|\frac{\left[u^2-w^2+x^2+4xy\right]^2+\left(4\pi{l}\Theta\right)^2}{\left[u^2-w^2+x^2-4xy\right]^2+\left(4\pi{l}\Theta\right)^2}\right|}\,,\label{qIETfinal5}\\
&\Psi(x,0)=-\frac{3}{4\Theta}\int_0^{\infty}\left\{\left[-\frac{B(u)}{\beta{U}(u)}+1\right]S(u)-\left[\frac{\Psi(u,0)}{\Phi(u,0)}\right]\left[S(u)-1\right]\right\}\frac{du}{u}\int_{|u-x|}^{u+x}w\left[S(w)-1\right]dw\nonumber\\&\qquad\quad\,\,\,\,\times\int_0^{\infty}dy\frac{y\exp{\left(\frac{y^2}{\Theta}-\bar{\mu}\right)}}{\left[\exp{\left(\frac{y^2}{\Theta}-\bar{\mu}\right)}+1\right]^2}\left\{\left[y^2-\frac{(u^2-w^2+x^2)^2}{16x^2}\right]\ln{\left|\frac{u^2-w^2+x^2+4xy}{u^2-w^2+x^2-4xy}\right|}+\frac{u^2-w^2+x^2}{2}\frac{y}{x}\right\},\label{qIETfinal6}\\
&S(x)=\frac{3}{2}\Theta\displaystyle\sum_{l=-\infty}^{\infty}\frac{\Phi(x,l)}{1+\displaystyle\frac{4}{\pi}\lambda{r}_{\mathrm{s}}\frac{1}{x^2}\left\{\left[1-\displaystyle\frac{{B}\left(x\right)}{\beta{U}(x)}\right]\Phi(x,l)-\Psi(x,l)\right\}}\,.\qquad\qquad\qquad\qquad\qquad\qquad\qquad\qquad\quad\,\,\,\label{qIETfinal7}
\end{align}
\end{widetext}
To speed-up the Matsubara series convergence, the non-interacting (Hartree Fock) SSF and the high-frequency short-wavelength limit of the auxiliary complex function $\Phi(x,l)$ are split-up from the sum\,\cite{TanIchi,IETChem,TanConv}. Summation up to $l=512$ suffices for convergence with the exception of highly degenerate strongly coupled states ($r_{\mathrm{s}}>100$, $\Theta<1$), for which summation up to $l=1024$ is necessary. To facilitate the short-wavelength convergence of the dynamic LFC of the qIET-based scheme, the well-behaved dynamic LFC of the qSTLS scheme is also split-up from the integral. In fact, a similar split-up of the static LFC of the STLS scheme from the static LFC of the IET-based scheme has proven to be beneficial for convergence in our earlier investigations\,\cite{IETChem}. The Broyles technique of mixing iterates\,\cite{mixite1}, as known from the numerical treatment of integral equation theory approximations\,\cite{mixite2,mixite3,mixite4,mixite5}, was also necessary to speed up and sometimes even to ensure the convergence of the dynamic qIET-based LFC. All improper integrals were numerically evaluated with the doubly adaptive Clenshaw-Curtis quadrature method; $0.1k_{\mathrm{F}}$ grid resolution and a $40k_{\mathrm{F}}$ upper cutoff were employed. The IET-based solution served as a good initial guess.

\begin{figure*}
	\centering
	\includegraphics[width=7.00in]{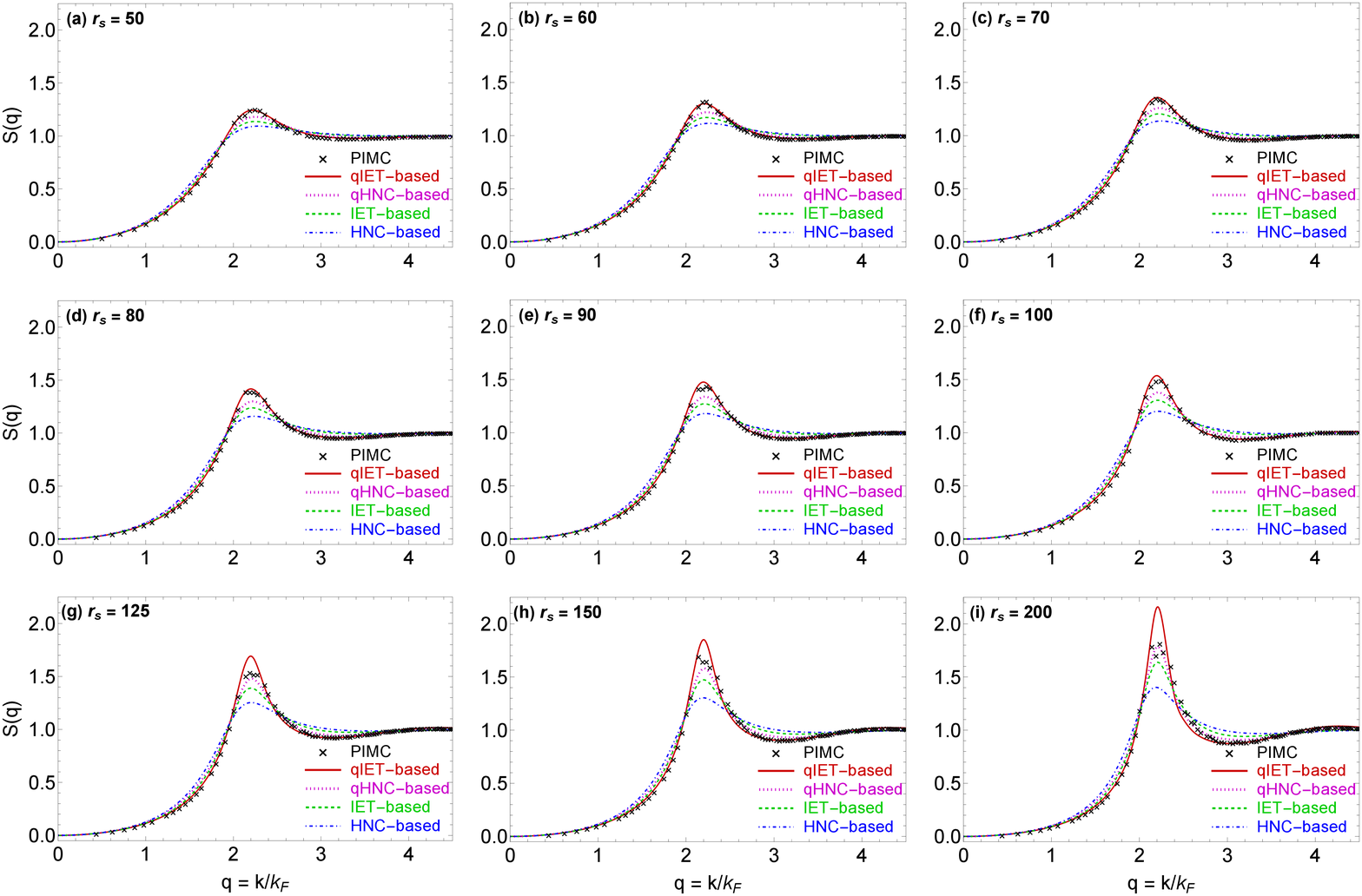}
	\caption{The electron liquid static structure factor for $\Theta=0.5\,\&\,r_{\mathrm{s}}=50-200$. Results from the HNC-based scheme (dot-dashed blue line), IET-based scheme (dashed green line), qHNC-based scheme (dotted magenta line), qIET-based scheme (solid red line) and PIMC simulations (black crosses). The proposed qIET-based scheme is near exact up to $r_{\mathrm{s}}=80$ and it remains very accurate up to $r_{\mathrm{s}}=100$. The proposed qHNC-based scheme becomes more accurate for $r_{\mathrm{s}}\gtrsim125$ due to a cancellation of errors.}\label{fig:rsdependence}
\end{figure*}

\begin{figure*}
	\centering
	\includegraphics[width=7.00in]{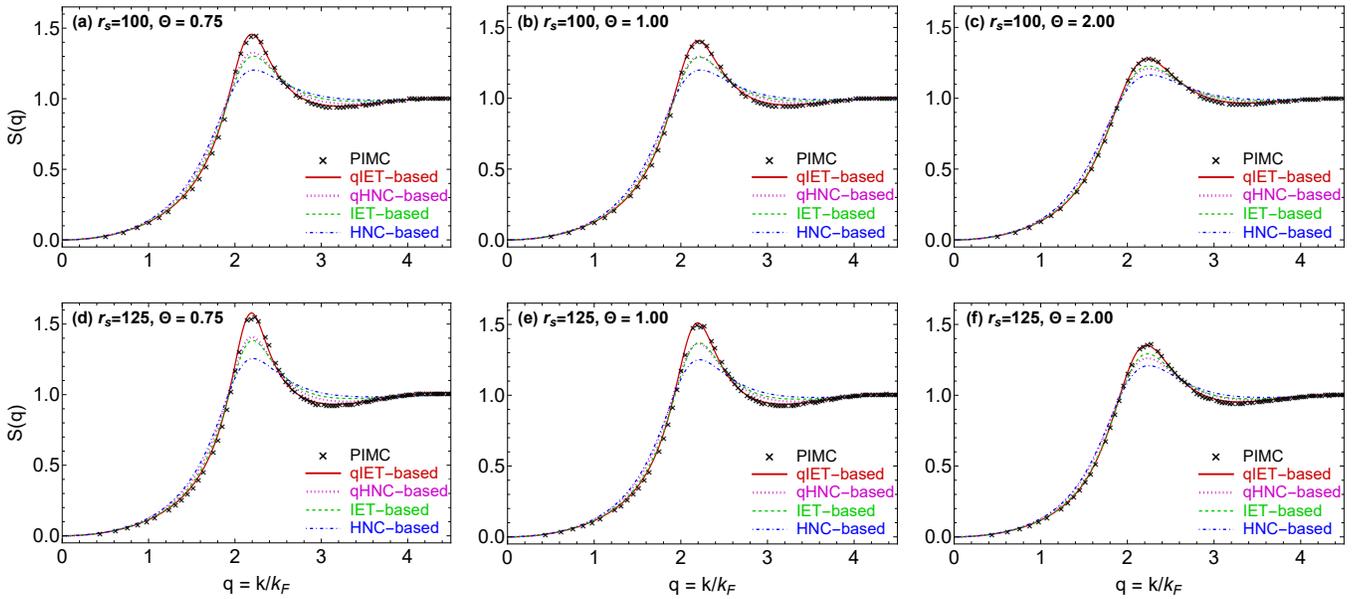}
	\caption{The electron liquid static structure factor for $r_{\mathrm{s}}=100,125\,\&\,\Theta=0.75,1.0,2.0$. Results from the HNC-based scheme (dot-dashed blue line), IET-based scheme (dashed green line), qHNC-based scheme (dotted magenta line), qIET-based scheme (solid red line) and PIMC simulations (black crosses). The qIET-based scheme yields near exact predictions at these six states.}\label{fig:thetadependence}
\end{figure*}

\emph{-- Results and comparison}. Characteristic SSFs as computed with the qIET-, qHNC-, IET-, HNC-based schemes and as extracted from PIMC simulations are illustrated in Figs.\ref{fig:rsdependence}\,\&\,\ref{fig:thetadependence}\,\cite{reposit}. The following robust conclusions have been drawn after the careful examination of the numerical results for all $20$ relevant state points: \textbf{(1)} The qIET scheme yields near exact SSFs for $r_{\mathrm{s}}\lesssim100$ and $\Theta\gtrsim0.5$. The SSFs remain very accurate up to $r_{\mathrm{s}}\lesssim150$, $\Theta\gtrsim0.75$, where observable deviations from the PIMC results begin to develop in the vicinity of the first SSF maximum. At even stronger coupling and even higher degeneracy, the qIET SSF maximum increasingly overshoots the PIMC SSF maximum. \textbf{(2)} For all states except those near the liquid-crystal boundary, the qIET scheme generates the most accurate SSFs across the entire wavenumber interval. This concerns the long wavelength range $k\lesssim2k_{\mathrm{F}}$, the near-Lorentzian region that surrounds the first maximum $k\sim2-2.5k_{\mathrm{F}}$ and the secondary extrema short wavelength region $k\gtrsim2.5k_{\mathrm{F}}$. \textbf{(3)} For states in the vicinity of Wigner crystallization (estimated by the onset of fluctuations near the PIMC SSF maximum), the qHNC-generated SSF features a more accurate Lorentzian region than the qIET-generated SSF, most probably due to a favorable cancellation between viscoelastic and bridge function effects. Even at the strongest coupling and the highest degeneracy considered, the qIET-generated SSFs are consistently the most accurate in the short and long wavelength ranges with the discrepancies from the PIMC SSF being nearly exclusively limited at the neighborhood of the global maximum. \textbf{(4)} The IET and the HNC SSFs always underestimate the PIMC SSFs near the maximum and always overestimate the PIMC SSFs within the short wavelength range, which leads to a favorable cancellation of errors in the computation of the interaction energy. This fortuitous characteristic is not shared by the qIET and the qHNC SSFs and translates to less accurate predictions for the interaction energy, see a detailed remark in what follows.\,\textbf{(5)} Regarding the static density response $\chi(\boldsymbol{k})$, the qIET scheme still yields by far the most accurate results. Overall, the agreement with PIMC simulations is less impressive. At the Fermi temperature $\Theta=1$, the qIET slightly undershoots the single extremum at moderate coupling $r_{\mathrm{s}}\sim50$ and accurately predicts the single extremum at strong coupling $r_{\mathrm{s}}\sim100-200$, but it no longer overlaps with the PIMC result in the short wavelength range. Note that the qHNC scheme yields a more accurate prediction for the single extremum only at $r_{\mathrm{s}}=200,\Theta=0.5$. The same conclusions naturally hold for the closely related static LFC, $G(\boldsymbol{k})$; $\omega=0$ in Eq.(\ref{densityresponseDLFC}). \textbf{(6)} As far as the pair correlation function is concerned, it is important to discuss the un-physical negative region near the origin, that is a common feature of all dielectric schemes\,\cite{DornRev,mapping}. This short range pathology is naturally a consequence of the approximate treatment of quantum effects, thus it is expected that the qIET- and the qHNC-based schemes will exhibit a better behavior than the IET- and the HNC-based schemes. In fact, while the radial extent of the negative region and the value of the on top pair correlation are similar for the four schemes, the negative contribution from the negative region is much smaller for the qIET- and the qHNC-based schemes. This is a consequence of the fact that qIET- \& qHNC-based pair correlation functions do not possess a short distance negative minimum in contrast to the IET- \& HNC-based pair correlation functions. This is hardly surprising, since this unphysical feature is inherited from the bare STLS scheme, for which it is omnipresent at strong coupling. \textbf{(7)} Concerning the route from structural to thermodynamic properties, the interaction energy per particle normalized by the Hartree energy $e^2/a_{\mathrm{B}}$ is given by\,\cite{DornRev,DornPoP}
\begin{align}
\widetilde{u}(r_{\mathrm{s}},\Theta)=\frac{1}{\pi\lambda{r}_{\mathrm{s}}}\int_0^{\infty}\left[S(x)-1\right]dx\,.\label{interactionenergies}
\end{align}
Unfortunately, the superiority of the qIET structural predictions does not translate to a superiority in thermodynamic predictions, since the HNC- \& IET-based schemes are distinguished by consistently favorable cancellation of errors between the short wavelength range and intermediate wavelength range. This characteristic is not shared by the qIET-generated SSFs whose small discrepancies with the PIMC-generated SSF are nearly exclusively concentrated in intermediate wavelengths. To be more specific, the qIET interaction energies exhibit an excellent agreement with the PIMC interaction energies having an accuracy within $1.53\%$ and an average accuracy of $0.89\%$. This pales in comparison to the remarkable performance of the IET interaction energies that have accuracy within $0.68\%$ and an average accuracy of $0.29\%$. Actually, it is comparable to the HNC interaction energies that are accurate within $1.37\%$ with a $0.96\%$ average accuracy. For completeness, note that qHNC interaction energies are accurate within $1.99\%$ with a $1.55\%$ average accuracy.

\emph{-- Possibility for further improvements.} There are two main drawbacks of the qIET-based scheme. \emph{First}, similar to the bare IET-based scheme, the classical Coulomb bridge function is introduced as a closed form $b(r,\Gamma)$ parametrization and not as a $b[g]$ functional\,\cite{OCPbrid,YOCPext}. This implies that the bridge function does not properly react to pure quantum mechanical effects and this also necessitates the introduction of an algebraic mapping between the classical OCP states ($\Gamma$) and the quantum OCP states ($r_{\mathrm{s}},\Theta$). The primitive mapping $\Gamma=2\lambda^2(r_{\mathrm{s}}/\Theta)$ employed is clearly insufficient, since it lacks a ground state limit, does not consider the Fermi energy contribution and does not depend on the spin polarization. Different strategies on the optimization of this mapping based on enforcing self-consistency (thermodynamic sum rules or frequency moment sum rules) will be explored in future studies. The above shortcoming manifests itself in the deviations from the exact results that arise at very low degeneracy parameters. \emph{Second}, the high temperature limit of the dynamic LFC of the qIET-based scheme exhibits a weak frequency dependence, \emph{i.e.} it is essentially static. Investigations of the classical OCP have demonstrated that a static LFC is accurate for moderate coupling but not in the vicinity of crystallization\,\cite{LFCOCP1,LFCOCP2}. The solution might lie in successfully interpolating between the static LFC limit and the exact high frequency LFC limit; the latter being tightly connected with the third frequency moment sum rule\,\cite{LFCOCP3,LFCOCP4,LFCOCP5}. The above shortcoming manifests itself in the deviations from the exact results that arise at very high quantum coupling parameters.

\emph{-- Summary and conclusions.} Our novel dielectric formalism scheme is tailor made for the strongly coupled regime of the finite temperature uniform electron fluid. Essentially, the scheme combines the interplay of quantum mechanical effects and thermal excitations as approximated in the qSTLS scheme with the interplay of strong correlations and thermal excitations as exactly described in the integral equation theory of liquids. The mathematical correspondence between the STLS and the qSTLS schemes is adopted as a general recipe to appropriately quantize semi-classical dielectric schemes. The physics approximations behind our scheme are not only uncontrollable but also rather obscured behind the mathematics. It is worth emphasizing that the recipe has been guided by available exact PIMC results. The truly remarkable agreement with the PIMC static structure factors in nearly the entire strongly coupled regime (with the exception of the Wigner crystallization vicinity) is a testament to the potential of the recipe. Future work will focus on exploring the dynamic structure factor predictions of the qIET scheme in search of exotic collective behavior and on improving the accuracy of the qIET scheme especially near the liquid-crystal phase boundary.

\emph{Acknowledgments.} This work was partly funded by the Swedish National Space Agency under grant no.\,143/16. This work was also partially supported by the Center for Advanced Systems Understanding (CASUS) which is financed by Germany's Federal Ministry of Education and Research (BMBF) and the Saxon state government out of the state budget that is approved by the Saxon State Parliament. The PIMC simulations were partly carried out at the Norddeutscher Verbund f\"ur Hoch- und H\"ochstleistungsrechnen (HLRN) under grant shp00026, and on a Bull Cluster at the Center for Information Services and High Performance Computing (ZIH) at Technische Universit\"at Dresden. The dielectric schemes were numerically solved on resources provided by the Swedish National Infrastructure for Computing (SNIC) at the NSC (Link{\"o}ping University) that is partially funded by the Swedish Research Council under grant agreement no.\,2018-05973.

\end{document}